\documentclass[]{WileyMSP-template}

\usepackage{amsmath}
\usepackage{amssymb}
\usepackage{amsfonts}
\usepackage{isomath}
\usepackage{tablefootnote}
\usepackage{graphicx}
\usepackage{color}
\usepackage[dvipsnames]{xcolor}
\usepackage{bbm}
\usepackage[breaklinks=true,colorlinks=true,linkcolor=blue,urlcolor=blue,citecolor=blue]{hyperref}
\usepackage{array}
\usepackage{booktabs}
\usepackage{mathptmx}
\usepackage{placeins}

\begin{document}

\pagestyle{fancy}
\rhead{\includegraphics[width=2.5cm]{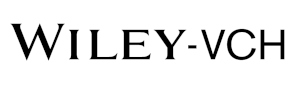}}

\title{DC-driven separation of fractional flux quanta in two-band superconductors}

\maketitle


\author{Anton O. Pokusinskyi*}
\author{Oleksandr V. Dobrovolskiy}


\dedication{}

\begin{affiliations}
A. O. Pokusinskyi, O. V. Dobrovolskiy\\
Cryogenic Quantum Electronics, Institute for Electrical Measurement Science and Fundamental Electrical Engineering\\ 
Technische Universit\"at Braunschweig\\ 
Hans-Sommer-Str. 66, 38106 Braunschweig, Germany\\
Email Address: anton.pokusinskyi@tu-braunschweig.de

A. O. Pokusinskyi, O. V. Dobrovolskiy\\
Laboratory for Emerging Nanometrology\\ 
Technische Universit\"at Braunschweig\\ 
Langer Kamp 6A-B, 38106 Braunschweig,  Germany\\

\end{affiliations}


\keywords{two-band superconductivity, fractional vortices, pinning, vortex matter}

\begin{abstract}

Two-band superconductors host vortices from superfluid condensates of different electron bands. These vortices carry a fractional flux quantum and attract each other, coalescing to form a composite vortex with the whole flux quantum $\phi_0$. However, due to the differences in viscosity and flux of the vortices across different bands, composite vortices may dissociate into fractional components. Here, we theoretically explore an approach to control the dissociation of composite vortices into fractional components and their separation into stationary and fast-moving ones through dc current and pinning strength variation. To this end, we numerically solve the dynamic equation of motion for a single dc-driven composite vortex in a periodic pinning potential. As the pinning strength increases, we observe a transition from depinning followed by dissociation in the weak-pinning regime to  dissociation from the pinned state in the strong-pinning regime. Under moderately strong pinning, fractional vortices from one condensate may become immobile while those from the other may even move faster than the original composite vortex just before the dissociation. The predicted pinning- and dc-controlled separation of fractional flux quanta appeals for experimental investigation and potential application in fluxonic devices. 

\end{abstract}


\section{Introduction}

The superconducting (SC) state exhibits a multicomponent nature in many materials, such as borocarbides\,\cite{Shu98prl}, NbSe$_2$\,\cite{Boa03prl},  MgB$_2$\,\cite{Mos09prl}, cuprates\,\cite{Hun08nat}, heavy fermion compounds\,\cite{Jou04prl,Sey05prl}, and iron-pnictides\,\cite{Son21nat,Zhe24arx,Igu23sci}.
In these materials, the SC state is formed by the electron condensates from different electron bands, and it can be described by multiple coupled order parameters\,\cite{Zeh13sst}. Distinct from conventional single-band SCs, multi-band SCs feature an unconventional structure of the mixed state\,\cite{Bab02prl} in which the interaction between vortices of equal vorticity is not necessarily purely repulsive, but in some cases may be characterized by long-range attraction\,\cite{Lin13prl}. A hallmark of multi-band SCs is the possibility for vortices to bear an arbitrary fraction of magnetic flux quantum, as predicted theoretically by Babaev\,\cite{Bab02prl} in the framework of the extended Ginzburg-Landau (GL) theory. The possibility of topological defects carrying fractional flux quanta\,\cite{Sil11prb,Lin14pcm} has garnered significant research interest, particularly due to unconventional pairing symmetries\,\cite{Iva01prl} and their potential applicability in topological quantum computing\,\cite{Nay08rmp}.

When the condensates have different phase rigidities, the vortices from each condensate might behave independently. Thus, an increase in temperature to the SC transition temperature $T_c$ was predicted to lead to the separation of fractional vortices, due to the differing SC gaps and phase fluctuations in each electron band\,\cite{Smo05prl}. Furthermore, with the aid of GL modeling, fractional vortices have been predicted to exist in mesoscopic two-component SCs\,\cite{Chi07epl,Geu10prb}. Experimentally, fractionalization of the quantum vortex core in KFe$_2$As$_2$ was recently observed by scanning tunneling microscopy\,\cite{Zhe24arx}. The mobility and manipulability of vortices carrying a temperature-dependent fraction of the flux quantum in Ba$_{1-x}$K$_x$Fe$_2$As$_2$ ($x\approx 0.77$) was demonstrated by scanning SC quantum interference device (SQUID) magnetometry\,\cite{Igu23sci}. 

Here, we propose a method to control composite vortex dissociation into fractional components, and their separation into pinned and moving ones, by adjusting dc current and pinning strength. The approach relies on the disparity in the viscosity and driving/pinning forces for fractional vortices in different bands, enabling the dissociation of composite vortices into fractional components. The theoretical analysis is based on solving the dynamic equation numerically for a single composite vortex driven by a dc current in a periodic pinning potential. The predicted effects call for examination in two-band SC thin films via vortex imaging\,\cite{Emb17nac,Igu23sci} and measurements of their current-voltage curves\,\cite{Gri18nsr,Bez22prb} at low magnetic fields.




\section{Model}

In the framework of the GL theory, the SC state in a two-band SC can be described by two weakly coupled order parameters $\Psi_i(r)$\,\cite{Zhi04prb,Lin13prl,Tan15sst}. Here, $i=1,2$ is the index of the wave functions of the SC condensates in each of the two electron bands, described by different values of the coherence length $\xi_i$ and the London penetration depth $\lambda_i$
\begin{equation}
\label{eq:eq_xi_i_lambda_i}
    \xi_i = 
    \sqrt{\frac{\hslash^{2}}{2 m_i |\alpha_i|}}
    , \qquad
    \lambda_i = 
    \sqrt{\frac{m_i}{2 \mu_{0} e^{2} 
    |\Psi_i|^{2}}},
\end{equation}
where $m_i$ is the effective electron mass and  $\alpha_i$ the GL parameter for the $i$-th condensate, $\hslash$ the reduced Planck constant, $e$ the electron charge, and $\mu_{0}$ the vacuum magnetic permeability.

When both condensates satisfy the condition for a vortex solution (GL parameter $\kappa_i = \lambda_i/\xi_i > 1/\sqrt{2}$), a two-band SC allows a moderate magnetic field to penetrate as fractional vortices, each carrying a fractional flux quantum \cite{Bab02prl,Lin13prl}
\begin{equation}
    \phi_i = 
    \phi_{0} \frac{\lambda_i^{-2}}{\lambda_{1}^{-2} + 
    \lambda_{2}^{-2}} = 
    \phi_{0} \frac{\lambda^{2}}{\lambda_i^{2}}
    ,
    \label{eq:eq_phi_i}
\end{equation}
where $\lambda^{-2} = \sum_{i = 1, 2} \lambda_i^{-2}$ is the effective penetration depth. 

\begin{figure}[t!]
    \centering
    \includegraphics[width=6.7cm]{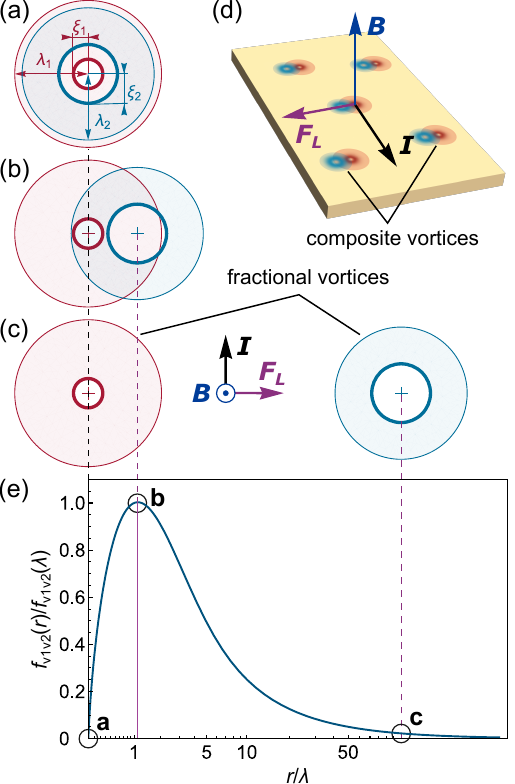}
    \caption{Relative positions of fractional vortex components for a composite vortex in the absence of a transport current (a), at the dissociation transition (b), and for the separated fractional vortices (c).
    (d) Geometrical model of a two-band SC film in a perpendicular magnetic field producing flux density $\mathbfit{B}$, where vortices experience a driving force $\mathbfit{F}_L$ induced by a dc transport current $\mathbfit{I}$. 
    (e) Dependence of the attraction force between the fractional vortices from different condensates on the distance between them by Equation\,(\ref{eq:eq_f_v1v2}), with the indicated relation to their relative positions in panels (a), (b), and (c).}
    \label{fig:img1_f_v1v2}
\end{figure}

Fractional vortices within the same condensate repel each other due to interactions in the condensate's quantum field. The net interaction between vortices from different condensates is attractive due to their coupling to the same vector potential field and direct interband coupling\,\cite{Sil11prb,Lin14pcm}. While the intraband vortex repulsion is long-ranged, the interband vortex attraction dominates at distances of the order of $\lambda$ and leads to the formation of composite vortices with flux $\phi_{1} + \phi_{2} = \phi_{0}$, see Figure\,\ref{fig:img1_f_v1v2}. The interband vortex attraction can be described by the force per unit vortex length \cite{Lin13prl, Pok24fnt}
\begin{equation}
    f_{v_1 v_2}(r_{1 2}) = \dfrac{\phi_{1} \phi_{2}}{2 \pi  \mu_{0} \lambda^{3}} 
        \text{Re} \left[\dfrac{\lambda}{r_{1 2}} - 
        K_{1}\left(\dfrac{r_{1 2}}{\lambda}\right)\right],
    \label{eq:eq_f_v1v2}
\end{equation}
where $\mathbfit{r}_{1 2} = \mathbfit{r}_{1} - 
\mathbfit{r}_{2}$ is the separation distance vector between the fractional components located at the points $\mathbfit{r}_i$ and $K_1$ is the Macdonald function of first order.

We consider a two-band SC film in a perpendicular magnetic field producing flux density $\mathbfit{B}$ and carrying a dc transport current $\mathbfit{I}$ of density $\mathbfit{j}$, see Figure\,\ref{fig:img1_f_v1v2}d. The film thickness $d$ is assumed to be smaller than the penetration depth, $d < \lambda$. This condition ensures a uniform distribution of the transport current across the film thickness, allowing the vortices to be treated as rigid cylindrical flux tubes, with their motion described by the coordinates of the vortex center.

In the flux flow state without defects, a vortex from the $i$-th condensate experiences the Lorentz force $f_{L,i} = j \phi_i$ induced by the transport current and the viscous force $f_i = \eta_iv_i$ due to the vortex motion through the SC. Here, $\eta_i = \phi_{0}^{2}/(2 \pi \xi_i^{2} \rho_{n})$ is the vortex viscosity, $\rho_{n}$ the normal resistivity, $v_i$ the vortex velocity, and the forces are expressed per unit vortex length. Accordingly, a composite vortex tends to dissociate into fractional components when the differences between $f_{L,i}$ and $\eta_{L,i}$ outweigh the interband vortex attraction. 

\begin{figure*}
    \centering
    \includegraphics[width=15cm]{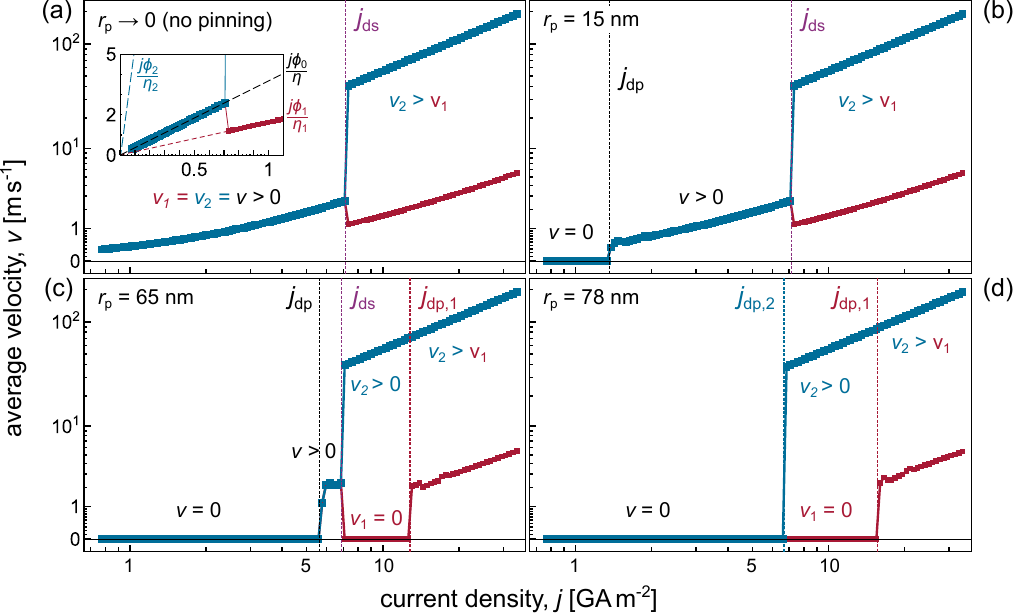}
    \caption{Dependence of the average velocity of fractional vortex components on the dc transport current density for a series of pinning strengths. $v$: velocity of the composite vortex; $v_i$: velocity of the fractional vortices from the $i$-th condensate; $j_{ds}$ current density of dissociation of the composite vortex into fractional components; $j_{dp}$: depinning current for the composite vortex; $j_{dp,i}$: depinning current for the fractional vortices from the $i$-th condensate. The inset in (a) shows the initial part of the $v(j)$ dependence with the respective asymptotics in linear scale.}
    \label{fig:img2_vortexVelocity}
\end{figure*}

The dependence of the interband vortex attraction force on the distance between fractional vortices from different condensates, calculated by Equation\,(\ref{eq:eq_f_v1v2}), is shown in Figure\,\ref{fig:img1_f_v1v2}e. It reveals three distinct regimes, as illustrated in Figure\,\ref{fig:img1_f_v1v2}a-c. Namely, in the absence of a transport current, the vortex cores are concentric, $|\mathbfit{r}_{1 2}|=0$, and there is no interaction force between the fractional components, Figure\,\ref{fig:img1_f_v1v2}a. As the transport current $j$  increases, $|\mathbfit{r}_{1 2}|$ increases and the fractional components begin to attract to one another. The attraction reaches a maximum at $|\mathbfit{r}_{1 2}| \approx \lambda$,  Figure\,\ref{fig:img1_f_v1v2}b. With a further increase of $j$, a certain current density is reached, which is known as the dissociation current $j_{ds}$. Indeed, for $|\mathbfit{r}_{1 2}| > \lambda$ the attraction between the fractional components weakens, leading to the dissociation of the composite vortex into fractional components, Figure\,\ref{fig:img1_f_v1v2}c. In this way, in the presence of a dc transport current, as long as 
$|\mathbfit{r}_{1 2}| \leq \lambda$, the fractional components remain bound and form a composite vortex whose motion can be described with the velocity $v = v_1 = v_2$ and the effective viscosity 
$\eta = \eta_1 + \eta_2$ \cite{Pok24fnt}. By contrast, when $|\mathbfit{r}_{1 2}| > \lambda$, the composite vortex dissociates into two fractional components which move with different velocities $v_1 \neq v_2$, due to the disparity in the viscosity $\eta_1 \neq \eta_2$ and the driving forces $f_{L,1} \neq f_{L,2}$.

In real materials, structural imperfections lead to vortex pinning, which can be modeled using a pinning potential with troughs described, e.g., by the Lorentzian \cite{Bla94rmp, Pas02ltp, Kla01prb, Pok24fnt}. Experimentally, such pinning sites can be realized, for instance, by linear columnar defects---radiation tracks \cite{Kal23sst}. In our model, the pinning force per unit length of vortex is described by the expression \cite{Pok24fnt}
\begin{equation}
    f_{p, i}(R_i) = 
    \alpha_{L, i}  
    \frac{R_i}{\left(1 + R_i^{2}/ 
    r_{p, i}^{2} \right)^{2}},
    \label{eq:eq_fp}
\end{equation}
where $\alpha_{L, i} \approx \eta_i \omega_{p}$ is the Labush parameter\,\cite{Pok22jap} with $\omega_{p}$ being the depinning frequency\,\cite{Lan97book,Pom08prb,Dob20pra}, $R_i$ the distance between the pinning site and the vortex from the $i$-th condensate, and $r_{p, i}$ the radius of the pinning potential trough. Equation\,\eqref{eq:eq_fp} allows for introducing a periodic pinning force $f_{pp,i}(x) = \sum_{i_{p}} f_{p, i}(x - d_{p} i_{p})$, where $i_{p}$ is the pinning site index and $d_{p}$ is the distance between two neighboring pinning sites.

By including all the afore-introduced forces, we obtain a system of dynamic equations governing the dc-driven motion of a single composite vortex in a periodic pinning potential. The single-vortex approximation is justified at fields $B \lesssim 2 B_{c1}$, where $B_{c1}$ is the lower critical field \cite{Bra91prl}. The obtained system of equations describes the time evolution of the coordinates $x_i(t)$ of the $i$-th fractional vortex components
\begin{equation}
    \begin{cases}
        \eta_{1} \dot{x}_{1} +
        f_{pp, 1}(x_{1}) +
        f_{v_1 v_2}(x_{1} - x_{2}) =
        j \phi_{1} \\[2mm]
        \eta_{2} \dot{x}_{2} +
        f_{pp, 2}(x_{2}) +
        f_{v_1 v_2}(x_{2} - x_{1}) =
        j \phi_{2} 
    \end{cases}.
    \label{eq:eq_composite_v_dynamics}
\end{equation}

The system of equations \eqref{eq:eq_composite_v_dynamics} is solved numerically for the dc transport current density $j$ varied between $0.1$ and $10^2$\,GA\,m$^{-2}$ and for pinning strengths tuned by varying the parameter $r_{p}$ between $0$ and $78$\,nm. The calculations were performed using the following parameters:
$\lambda_{1} = 290$\,nm, 
$\lambda_{2} = 250$\,nm,
$\xi_{1} = 10$\,nm, 
$\xi_{2} = 50$\,nm, 
and $\rho_{n} = 0.1\,\mu\mathrm{\Omega}$cm. While these values are not specific to a particular two-band SC material, they fall within a comparable range for FeSe-based films\, \cite{Pur19pt,Lin15sst}. An estimate for the depinning frequency $\omega_{p} = 100$\,MHz \cite{Lia18pra, Pom20sst} was used to calculate the Labusch parameter $\alpha_{L, i} \approx \eta_i \omega_{p}$. The solution of the system of equations \eqref{eq:eq_composite_v_dynamics} is discussed next.

\section{Results}

Figure\,\ref{fig:img2_vortexVelocity} presents the calculated current--velocity $v(j)$ characteristics of the system under study for a series of pinning strengths $r_p$, where $v$ is the average velocity of the fractional vortex components and $j$ is the dc transport current density.

We begin our analysis by considering the case without pinning, as shown in Figure\,\ref{fig:img2_vortexVelocity}a. At weak and moderately strong currents $j< j_{ds}$, $v(j)$ increases linearly with increasing $j$. This means that both components form a composite vortex moving with $v_1 = v_2 = v = \phi_{0} j / (\eta_{1} + \eta_{2})$. As soon as $j$ reaches $j_{ds}$, the $v(j)$ line abruptly splits into two branches with velocities $v_{i} = \phi_i j / \eta_i$. Herewith, one of the fractional components acquires a notably higher velocity than the composite vortex had just before the dissociation transition, $v_2=j\phi_2/\eta_2 > v=j\phi_0/\eta$, see the inset in Figure\,\ref{fig:img2_vortexVelocity}a. The other fractional component begins to move with $v_1 = j\phi_1/\eta_1 < v$, that is first even slower than the composite vortex at $j \lesssim j_{ds}$. With a further increase of $j$, both $v_1$ and $v_2$ increase.

The introduction of weak pinning results in the emergence of a zero-velocity plateau in the $v(j)$ curve at small $j$, see Figure\,\ref{fig:img2_vortexVelocity}b, indicating a regime where vortices remain immobile despite the applied current. Indeed, a composite vortex remains pinned for currents smaller than the depinning current, $j < j_{dp}$. At larger $j > j_{dp}$, the $v(j)$ curve qualitatively mimics its behavior in the absence of pinning. The little ripple in $v(j)$ at $j\gtrsim j_{dp}$ is attributed to the poor averaging of the vortex velocity when the slow vortex motion in the periodic pinning potential entails a strong anharmonicity. 

A further increase in the pinning strength introduces a qualitatively new feature in the $v(j)$ curve, see Figure\,\ref{fig:img2_vortexVelocity}c. Namely, at $j= j_{ds}$, a composite vortex is split into a fractional component with larger core radius $\xi_2$, whose velocity increases drastically, $v_2 \gg v$, and a fractional component with smaller core radius $\xi_1$, whose velocity $v_1$ drops down to zero. We attribute the ``re-pinning'' of the fractional component with smaller $\xi_1$ to its higher pinning sensitivity, as is also known, e.g., for high-$T_c$ cuprates with very small coherence length\,\cite{Bla94rmp}. With a further increase of $j$ to $j_{dp,1}$, the ``slow'' fractional component is depinned once again, such that $v_2 > v_1 > 0$. 

Finally, with a further increase in the pinning strength, one can realize a regime when $j_{dp} \geq j_{ds}$, see Figure\,\ref{fig:img2_vortexVelocity}d. In this case, the component with higher mobility and larger vortex core is depinned at $j_{dp,2}$ and this depinning is preceded by a dissociation of the composite vortex in the pinned state. The other, less mobile fractional component with smaller vortex core remains pinned to the defect as long as $j< j_{dp,1}$. In this way, the significant disparity between the depinning current densities $j_{dp,2}$ and $j_{dp,1}$ for the fractional components provides a pathway for their separation in two-band SCs. 

\section{Discussion}

\begin{figure}[t!]
    \centering
    \includegraphics[width=8cm]{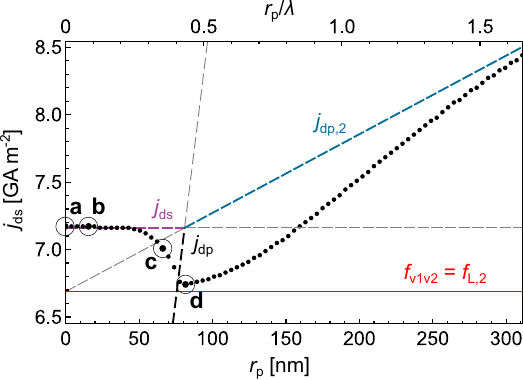}
    \caption{Dependence of the composite vortex dissociation current $j_{ds}$ on the parameter $r_{p}$ describing the pinning strength. Circles indicate the $r_{p}$ values which correspond to the respective panels in Figure\,\ref{fig:img2_vortexVelocity}.
    Straight lines denote the following asymptotics: 
    $f_{v_{1} v_{2}}$: maximum of the interband attraction force by Equation\,\eqref{eq:eq_f_v1v2} between the fractional components;
    $j_{ds}$: dissociation current density in the case of weak pinning;
    $j_{dp}$: depinning current density for the composite vortex;
    $j_{dp,2}$: depinning current density for the more mobile fractional component;
    $f_{L,2}$: Lorenz force for the more mobile fractional component.}
    \label{fig:img3_j_ds}
\end{figure}

We proceed to a discussion of the evolution of the depinning and dissociation currents with increasing pinning strength. Usually, an increase in the pinning strength impedes the vortex motion and leads to an enhancement of the depinning current. However, in the studied system, an increase in the pinning strength can stipulate a mobile vortex state at smaller $j$ than at lower pinning intensities. Namely, while pinning is weak (points \textbf{a} and \textbf{b} in Figure\,\ref{fig:img3_j_ds}), the dissociation current $j_{ds}$ remains constant and equal to the asymptotic $j_{ds}$ value given by the dissociation current for a dc-driven composite vortex without pinning. 

With increasing pinning intensity (point \textbf{c} in Figure\,\ref{fig:img3_j_ds}), the increasing $j_{dp}$ approaches the $j_{ds}$ asymptotics and an interesting regime emerges when the size of the pinning potential wells falls within the range $2\xi_1 < r_p \lesssim 2\xi_2$. Note that $r_p$ characterizes the spatial scale where the pinning forces are actually acting, which due to the Lorentzian shape differs from $2r_p$. Accordingly, when $50$\,nm$<r_p<80$\,nm, the potential trough primarily pins the vortices with the smaller cores $\sim2\xi_1$ while the pinning potential width is not enough to efficiently pin the vortices with the larger cores $\sim2\xi_2$. Consequently, the pinning enhances the disparity between the effective driving forces for different components and this leads to a decrease in the dissociation current (point \textbf{c} in Figure\,\ref{fig:img3_j_ds}).
Indeed, the ratio of the pinning forces, $\alpha_{L, 1}/\alpha_{L, 2}$, is about $25$. Hence, during the composite vortex depinning process, one fractional component can detach from the pinning site earlier, as soon as the condition $f_{v_{1} v_{2}} = f_{p, 1}$ is satisfied for dissociation. This occurs when the asymptotic $j_{dp}$ crosses the line $f_{v_{1} v_{2}} = f_{L, 2}$ in Figure\,\ref{fig:img3_j_ds}.

As the pinning strength increases further, $j_{dp}$ reaches $j_{ds}$ and the dependence $j_{ds}(r_p)$ goes through a minimum (point \textbf{d} in Figure\,\ref{fig:img3_j_ds}) and then increases again. This behavior can be explained by the consideration that, as $r_p$ increases, the pinning potential well becomes wide enough to pin both fractional components, thereby diminishing the disparity in the effective driving forces. Simultaneously, when the asymptotic $j_{ds}$ is crossed by $j_{dp}$, the process switches from depinning followed by dissociation to a consecutive depinning of the fractional components. As 
$r_p$ increases further, 
$j_{ds}$ approaches the asymptotic value $j_{dp, 2}$, which corresponds to the depinning current of the more mobile fractional component.

The minimum in $j_{ds}$ at point (d) in Figure\,\ref{fig:img3_j_ds} can thus be explained by a combined effect of the spatial anchoring of the less-mobile fractional vortices in the pinning potential trough and the negligible pinning for the more mobile fractional vortices. As a result, at $j_{ds} = j_{dp}$, the numerically obtained data (dotted curve in Figure\,\ref{fig:img3_j_ds}) almost reach the theoretically minimal value of the transport current required for the composite vortex dissociation, which is obtained assuming that one of the fractional components remains stationary, making it easier to separate the fractional components by pulling on only one of them, $f_{v_{1} v_{2}} = f_{L, i}$.

The explanation of the revealed $j_{ds}$ minimum can further be supported by the consideration of the maximum distances at which the fractional components can separate from each other while moving with the same velocity $v$ as a composite vortex. Namely, in the process of composite vortex depinning, the distances between its fractional components can become much larger (up to $\lambda$) than $r_p$, see Figure\,\ref{fig:img1_f_v1v2}.  The effect stays noticeable as long as $r_p$ remains smaller than the distance between the fractional components, $r_{p} < \lambda$. At larger distances between the fractional components, $r_{p} \gtrsim \lambda$, the numerically obtained dependence $j_{ds}(r_p)$ tends to the $j_{dp, 2}$ asymptotic. 

We would like to complete our discussion by commenting on possible experimental examinations of the predicted effects through electrical resistance measurements. First, since the experimental observable in such experiments is the electrical voltage, the calculated vortex velocities need to be multiplied with the respective flux fractions given by Equation\,\eqref{eq:eq_phi_i}. Second, the transport currents corresponding to the regime $v_1 =0$ and $v_2>0$ in Figure\,\ref{fig:img2_vortexVelocity}c should be smaller not only than the GL depairing current but also than the flux-flow instability current\,\cite{Dob24inb}. Close to $T_c$, the flux-flow instability causes a breakdown of the low-resistive state at high vortex velocities due to non-equilibrium effects affecting the electron distribution function via the Larkin-Ovchinnikov mechanism\,\cite{Lar75etp}. At low temperatures $T \ll T_c$, the instability is usually of thermal nature and described by the Kunchur mechanism\,\cite{Kun02prl}. In general, the flux-flow instability can be suppressed via collective ordering effects by exploiting, e.g., the vortex guiding effect\,\cite{Dob20pra}, edge-quality improvement\,\cite{Bud22pra}, and covering the SC with a capping layer\,\cite{Att12pcm}. The vortex velocities of interest (below $100$\,m\,s$^{-1}$ in Figure\,\ref{fig:img1_f_v1v2}) can be tuned by such material characteristics as, e.g., the normal-state resistivity. Such velocities appear attainable, given that vortex velocities of up to $2$\,km\,s$^{-1}$ have been observed experimentally for layered Fe(SeTe)\,\cite{Gri18nsr}.

\section{Conclusion}

We have theoretically studied the dynamics of a dc-driven composite vortex in two-band superconductors with a periodic pinning potential. By varying the dc transport current and pinning strength, we observed a branching in the current--velocity characteristics, resulting from the dissociation of the composite vortex into fractional components moving with different velocities. This disparity arises from the differing superconducting parameters of each condensate. Under certain conditions, we showed that the velocity of one component can drop to zero within a specific range of dc currents, while the velocity of the other component can significantly increase compared to the original composite vortex. The disparity between the depinning currents for the fractional components provides a pathway for their separation in two-band superconductors. The obtained results appeal for experimental examination through vortex imaging and measurements of
the current-voltage curves at low magnetic fields.




\medskip
\textbf{Acknowledgements} \par 
The authors thank A.L. Kasatkin for fruitful discussions. The work of A.P. was funded by the Deutsche Forschungsgemeinschaft (DFG, German Research Foundation) under Germany’s Excellence Strategy -- EXC-2123 QuantumFrontiers -- 390837967. A.P. gratefully acknowledges the use of the CryoCore simulation workstation at CryoQuant/TU Braunschweig. The research is based upon work from COST Action CA21144 (SuperQuMap) supported by the European Cooperation in Science and
Technology.

\medskip
\textbf{Conﬂict of Interest} \par
The authors declare no conﬂict of interest.

\medskip
\textbf{Data Availability Statement} \par
The data underlying this study are openly available in Mendeley Data at doi: 10.17632/2cb9pf4nfz.

%
\bibliographystyle{MSP}
\bibliography{main.bib}

\end{document}